\documentclass[twocolumn,prl,eqsecnum,aps,floats]{revtex4}
\usepackage{dcolumn}
\usepackage{amsmath}
\usepackage{graphicx}
\usepackage{rotating}
\usepackage{amsfonts}
\usepackage{amssymb}
\usepackage[hang]{subfigure}
\begin{document}
% LIST_OF_AUTHORS_R1.TEX                 12/5/02            
%
\author{                                                                      
%% names begin here                                                           
V.M.~Abazov,$^{21}$                                                           
B.~Abbott,$^{55}$                                                             
A.~Abdesselam,$^{11}$                                                         
M.~Abolins,$^{48}$                                                            
V.~Abramov,$^{24}$                                                            
B.S.~Acharya,$^{17}$                                                          
D.L.~Adams,$^{53}$                                                            
M.~Adams,$^{35}$                                                              
S.N.~Ahmed,$^{20}$                                                            
G.D.~Alexeev,$^{21}$
% ...
\\                                                                            
\vskip 0.30cm                                                                 
\centerline{(D\O\ Collaboration)}                                             
\vskip 0.30cm                                                                 
}                                                                             
\address{                                                                     
\centerline{$^{1}$Universidad de Buenos Aires, Buenos Aires, Argentina}       
\centerline{$^{2}$LAFEX, Centro Brasileiro de Pesquisas F{\'\i}sicas,         
                  Rio de Janeiro, Brazil}                                     
\centerline{$^{3}$Universidade do Estado do Rio de Janeiro,                   
                  Rio de Janeiro, Brazil}                                     
\centerline{$^{4}$Institute of High Energy Physics, Beijing,                  
                  People's Republic of China}                                 
\centerline{$^{5}$Universidad de los Andes, Bogot\'{a}, Colombia}             
\centerline{$^{6}$Charles University, Center for Particle Physics,            
                  Prague, Czech Republic}                                     
\centerline{$^{7}$Institute of Physics, Academy of Sciences, Center           
                  for Particle Physics, Prague, Czech Republic}               
\centerline{$^{8}$Universidad San Francisco de Quito, Quito, Ecuador}         
\centerline{$^{9}$Institut des Sciences Nucl\'eaires, IN2P3-CNRS,             
                  Universite de Grenoble 1, Grenoble, France}                 
\centerline{$^{10}$CPPM, IN2P3-CNRS, Universit\'e de la M\'editerran\'ee,     
                  Marseille, France}                                          
\centerline{$^{11}$Laboratoire de l'Acc\'el\'erateur Lin\'eaire,              
                  IN2P3-CNRS, Orsay, France}                                  
\centerline{$^{12}$LPNHE, Universit\'es Paris VI and VII, IN2P3-CNRS,         
                  Paris, France}                                              
\centerline{$^{13}$DAPNIA/Service de Physique des Particules, CEA, Saclay,    
                  France}                                                     
\centerline{$^{14}$Universit{\"a}t Mainz, Institut f{\"u}r Physik,            
                  Mainz, Germany}                                             
\centerline{$^{15}$Panjab University, Chandigarh, India}                      
\centerline{$^{16}$Delhi University, Delhi, India}                            
\centerline{$^{17}$Tata Institute of Fundamental Research, Mumbai, India}     
\centerline{$^{18}$CINVESTAV, Mexico City, Mexico}                            
\centerline{$^{19}$FOM-Institute NIKHEF and University of                     
                  Amsterdam/NIKHEF, Amsterdam, The Netherlands}               
\centerline{$^{20}$University of Nijmegen/NIKHEF, Nijmegen, The               
                  Netherlands}                                                
\centerline{$^{21}$Joint Institute for Nuclear Research, Dubna, Russia}       
\centerline{$^{22}$Institute for Theoretical and Experimental Physics,        
                   Moscow, Russia}                                            
\centerline{$^{23}$Moscow State University, Moscow, Russia}                   
\centerline{$^{24}$Institute for High Energy Physics, Protvino, Russia}       
\centerline{$^{25}$Lancaster University, Lancaster, United Kingdom}           
\centerline{$^{26}$Imperial College, London, United Kingdom}                  
\centerline{$^{27}$University of Arizona, Tucson, Arizona 85721}              
\centerline{$^{28}$Lawrence Berkeley National Laboratory and University of    
                  California, Berkeley, California 94720}                     
\centerline{$^{29}$University of California, Davis, California 95616}         
\centerline{$^{30}$California State University, Fresno, California 93740}     
\centerline{$^{31}$University of California, Irvine, California 92697}        
\centerline{$^{32}$University of California, Riverside, California 92521}     
\centerline{$^{33}$Florida State University, Tallahassee, Florida 32306}      
\centerline{$^{34}$Fermi National Accelerator Laboratory, Batavia,            
                   Illinois 60510}                                            
\centerline{$^{35}$University of Illinois at Chicago, Chicago,                
                   Illinois 60607}                                            
\centerline{$^{36}$Northern Illinois University, DeKalb, Illinois 60115}      
\centerline{$^{37}$Northwestern University, Evanston, Illinois 60208}         
\centerline{$^{38}$Indiana University, Bloomington, Indiana 47405}            
\centerline{$^{39}$University of Notre Dame, Notre Dame, Indiana 46556}
\centerline{$^{@}$Now at California Institute of Technology, Pasadena, California 91125}       
\centerline{$^{40}$Iowa State University, Ames, Iowa 50011}                   
\centerline{$^{41}$University of Kansas, Lawrence, Kansas 66045}              
\centerline{$^{42}$Kansas State University, Manhattan, Kansas 66506}          
\centerline{$^{43}$Louisiana Tech University, Ruston, Louisiana 71272}        
\centerline{$^{44}$University of Maryland, College Park, Maryland 20742}      
\centerline{$^{45}$Boston University, Boston, Massachusetts 02215}            
\centerline{$^{46}$Northeastern University, Boston, Massachusetts 02115}      
\centerline{$^{47}$University of Michigan, Ann Arbor, Michigan 48109}         
\centerline{$^{48}$Michigan State University, East Lansing, Michigan 48824}   
\centerline{$^{49}$University of Nebraska, Lincoln, Nebraska 68588}           
\centerline{$^{50}$Columbia University, New York, New York 10027}             
\centerline{$^{51}$University of Rochester, Rochester, New York 14627}        
\centerline{$^{52}$State University of New York, Stony Brook,                 
                   New York 11794}                                            
\centerline{$^{53}$Brookhaven National Laboratory, Upton, New York 11973}     
\centerline{$^{54}$Langston University, Langston, Oklahoma 73050}             
\centerline{$^{55}$University of Oklahoma, Norman, Oklahoma 73019}            
\centerline{$^{56}$Brown University, Providence, Rhode Island 02912}          
\centerline{$^{57}$University of Texas, Arlington, Texas 76019}               
\centerline{$^{58}$Texas A\&M University, College Station, Texas 77843}       
\centerline{$^{59}$Rice University, Houston, Texas 77005}                     
\centerline{$^{60}$University of Virginia, Charlottesville, Virginia 22901}   
\centerline{$^{61}$University of Washington, Seattle, Washington 98195}       
}                                                                             
%end  

\def\MET{\mbox{${\hbox{$E$\kern-0.6em\lower-.1ex\hbox{/}}}_T$}}

%\begin{document}

\title {Search for Large Extra Dimensions in the Monojet + $\MET$ Channel at D\O}
%\maketitle
%\centerline{\paperversion, \today}
\begin{abstract}
We present a search for large extra dimensions (ED) in $p\bar{p}$ collisions at a center-of-mass energy of 1.8 TeV using data collected by the D\O\ detector at the Fermilab Tevatron in 1994-1996. Data corresponding to 78.8 $\pm$ 3.9 pb$^{-1}$ are examined for events with large missing transverse energy, one high-p$_T$ jet, and no isolated muons. There is no excess observed beyond expectation from the standard model, and we place lower limits on the fundamental Planck scale of 1.0 TeV and 0.6 TeV for 2 and 7 ED, respectively.
\end{abstract}
\maketitle
%\pacs{11.25.Wx  11.10.Kk  13.85.Rm}
%\twocolumn
The standard model (SM) of particle physics is a spectacular scientific achievement, with nearly every prediction confirmed to a high degree of precision. Nevertheless, the SM still has unresolved and unappealing characteristics, including the problem of a large hierarchy in the gauge forces, with gravity being a factor of $10^{33}$ -- $10^{38}$ weaker than the other three. A new framework for addressing the hierarchy problem was proposed recently by Arkani-Hamed, Dimopoulos, and Dvali \cite{add}, through the introduction of large compactified extra spatial dimensions in which only gravitons propagate. In the presence of $n$ of these extra dimensions, the fundamental Planck scale in $4+n$ dimensions can be lowered to the TeV range, i.e., to a value comparable to the scale that characterizes the other three forces, thereby eliminating the puzzling hierarchy.

The radius ($R$) of the compactified extra dimensions can be expressed as a function of a fundamental Planck scale, $M_D \approx 1$~TeV, the number
of extra dimensions $n$, and the usual Planck scale $M_{\rm Pl} = 1/\sqrt{G_N}$. Assuming compactification on a torus, the relationship is \cite{giudice}:
\[R=\frac{1}{\sqrt[ n]{8\pi}M_D}(M_{\rm Pl}/M_D)^{2/n}\text{.}\]
The value $n=1$ is ruled out by the $1/r^{2}$ dependence of the gravitational force at large distances. The current limits from tests of gravity at short distances \cite{adelwash}, as well as from stringent astrophysical and cosmological bounds \cite{astro}, have significantly constrained the case of two extra dimensions. For $n>$ 2, the constraints from direct gravitational measurements and cosmological observations are relatively weak. However, high-energy colliders can provide effective ways to test such models of large ED \cite{blackhole}.

In the framework of large ED, at high energies, the strength of gravity in four dimensions is enhanced through a large number of graviton excitations, or Kaluza-Klein modes $(G_{\rm KK})$ \cite{kk}. This leads to new phenomena predicted for collisions at high energy \cite{giudice,colliders}: virtual graviton exchange and direct graviton emission. Virtual graviton exchange leads to anomalous difermion and diboson production, and searches for these effects have been pursued at the Tevatron \cite{ledfermi}, LEP \cite{ledlep}, and HERA \cite{ledhera}. For real graviton emission, since the graviton escapes detection, the signature involves large missing transverse energy $\MET$, accompanying a single jet or a vector boson at large transverse momentum. LEP experiments~\cite{ledlep} and the CDF collaboration~\cite{ledcdf} have recently set limits on $M_D$ based on $\gamma+G_{\rm KK}$ production. 

In this Letter, we report results of the first search for large ED in the jet + $\MET$ channel. The advantage of this channel is its relatively large cross section, with the tradeoff of large background. Besides $Z(\nu\bar{\nu})~+~$jets, which is the irreducible background, there are instrumental backgrounds from mismeasurement of, e.g., jet $E_T$, vertex position, undetected leptons, cosmic rays, etc. The data used for this search were collected in 1994 -- 1996 by the D\O\ collaboration \cite{d0} at the Fermilab Tevatron, using proton-antiproton collisions at a center-of-mass energy of 1.8 TeV. This sample, representing an integrated luminosity of 78.8 $\pm$ 3.9 pb$^{-1}$, was obtained using $\MET$ triggers with thresholds between 35 and 50~GeV. 

The D\O\ detector \cite{d0} consists of three major components: an inner detector for tracking charged particles, a uranium/liquid-argon calorimeter for measuring electromagnetic and hadronic showers, and a muon spectrometer consisting of magnetized iron toroids and three layers of drift tubes. Jets are measured with an energy resolution of approximately $\sigma (E)/E$ = 0.8/$\sqrt{E}$ ($E$ in GeV). $\MET$ is measured with a resolution of $\sigma({\MET})$ = $a+b\times S_T+c\times S_T^{2}$, where $S_T$ is the scalar sum of transverse energies in all calorimenter cells, $a$ = 1.89 $\pm$ 0.05~GeV, $b$ = (6.7 $\pm$ 0.7) $\times$ 10$^{-3}$, and $c$ = (9.9 $\pm$ 2.1) $\times$ 10$^{-6}$~GeV$^{-1}$ \cite{metreso}.

After eliminating events of poor quality (e.g., containing hot cells
in the calorimeter), events with one central (detector pseudorapidity
$|\eta_{d}| \le 1.0$ \cite{etad}) high-$E_T$ jet ($j_1$) and large
$\MET$, with $E_T(j_1) > 150$ GeV and $\MET > 150$ GeV, were selected
for further study. Since signal can contain initial or final-state
radiation (ISR or FSR), additional jets can also be present in such
interactions. To improve signal efficiency, we therefore allow
additional jets in the event, but require the second jet ($j_2$) to
have $E_T(j_2)< 50$~GeV, which reduces the background from dijet
production, while retaining most of the signal containing ISR or
FSR. To suppress $W$ or $Z$ production with a muon in the final state,
as well as to reduce the background from cosmic rays, we reject events
with isolated muons, that is, with $\Delta \cal{R}$$(j_1, \mu) > 0.5$,
based mainly on information from the muon system (referred to at D\O\
as Isolated Muon Veto 1), and based on information from the
calorimeter (Isolated Muon Veto 2), to suppress $W$ or $Z$ production
with a muon in the final state as well as to reduce the background
from cosmic rays. (The separation between objects is defined as
$\Delta \cal{R}$~=~$\sqrt{(\Delta\eta)^{2}+(\Delta\phi)^2}$, where
$\eta$ is the pseudorapidity and $\phi$ is the azimuthal angle.)
Backgrounds with isolated electrons are expected to be small, and we
therefore do not use any special criteria to suppress electrons. We
also require $\Delta\phi(j_2, \MET)>15^{\circ}$, to reduce the
background from mismeasured jets in multijet (``QCD'') events. An
additional source of background is from hard bremsstrahlung of
cosmic-ray muons that pass through the D\O\ calorimeter. For any
showers induced by photons radiated in the hadronic layers of the
calorimeter, the resulting ``jets'' usually contain only a handful of
cells with significant energy deposition, and such jets therefore fail
our quality criteria. However, for bremsstrahlung that occurs in the
EM section of the calorimeter, the shower is usually reconstructed as
an EM object, and not as a jet. Thus, most of the background arises
from showers that originate near the regions of confusion at the
interface of the EM and hadronic calorimeters. To reduce this
background, we remove events with such ``jets'', as well as events
that contain ``tracks'' of minimum energy deposition, which are
typical of muons observed in the finely segmented D\O\
calorimeters. Jet ``pointing'', based on tracking information in the
leading jet ($j_1$), is used to confirm the longitudinal position of
the primary vertex by requiring that $\Delta
z(j_1$-vertex,~primary-vertex$) \le 10$~cm. This suppresses background
from cosmic rays as well as from events with incorrectly reconstructed
primary vertexes. The requirements on $\eta_{d}$ of the leading jet
and on the event primary vertex confirmation are chosen to maximize
the significance of signal relative to background. A total of 38
events remain in the data sample after applying all selections, as
shown in Table \ref{tb:tb1}.
\begin{table}[tb] 
\centering
\caption
{Observed number of events passing each requirement in the data with $E_T(j_1) > 150$~GeV, $\MET > 150$~GeV, and $E_T(j_2)< 50$~GeV.\label{tb:tb1}}
\begin{tabular}
[c]{|l|c|}\hline
Criterion & \multicolumn{1}{|c|}{Number of Events}\\\hline
Event Quality & 301,325\\\hline
Isolated Muon Veto 1  & 296,742\\\hline
Leading-jet, $\MET$ and & 141\\
Second-jet Requirement & \\\hline
$\Delta\phi(j_2, \MET)$ & 129\\\hline
Cosmic Ray Rejection & 69\\\hline
Primary Vertex Confirmation & 39\\\hline
Isolated Muon Veto 2 & 38\\\hline
\end{tabular}
\vspace{-5pt}
\end{table}

The {\sc PYTHIA} Monte Carlo (MC) generator \cite{pythia}, with implementation of the ED signal via Ref. \cite{matchev}, including the parton-level subprocesses $qg \to qG_{\rm KK},$ $q\bar q\to gG_{\rm KK},$ and $gg \to gG_{\rm KK}$, is used to generate signal events. This is followed by processing through D\O\ fast-detector simulation {\sc QSIM} routines\cite{qsim3587}. The signal is simulated for $n$ = 2 to $n$ = 7 extra dimensions, with $M_D$ ranging from 600~GeV to 1400~GeV in 200~GeV steps. The acceptance for signal varies from about 5$\%$ to 8$\%$, depending on the values of $n$ and $M_D$. The 13$\%$ contribution to the uncertainty on the overal acceptance is due to the limited size of the MC samples, and is of the same order as the contributions from the jet-energy scale \cite{jes} (5--12$\%$) and the choice of parton distribution functions (PDFs) (3--5$\%$). (The CTEQ3M set of parton distribution functions (PDFs) \cite{pdf} was used as a default choice in the analysis.)  

\begin{figure}[htb]
  \vbox{
  \vspace{-0.15in}
  \centering
	\subfigure{
	\label{fig:fig1a}
          \includegraphics
[width=3in]
{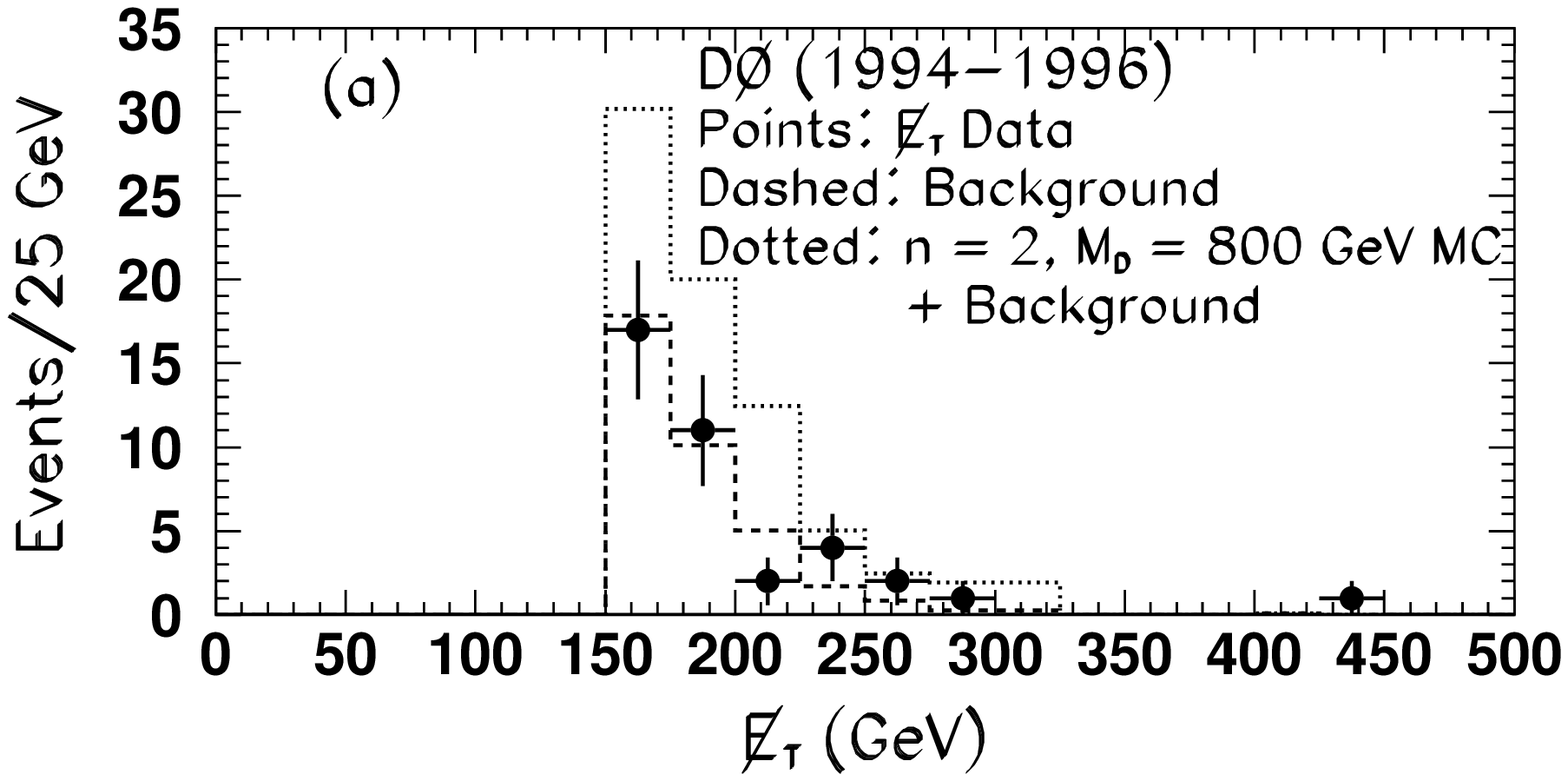}
	}
	\subfigure{
	\label{fig:fig1b}
	\includegraphics
[width=3in]
{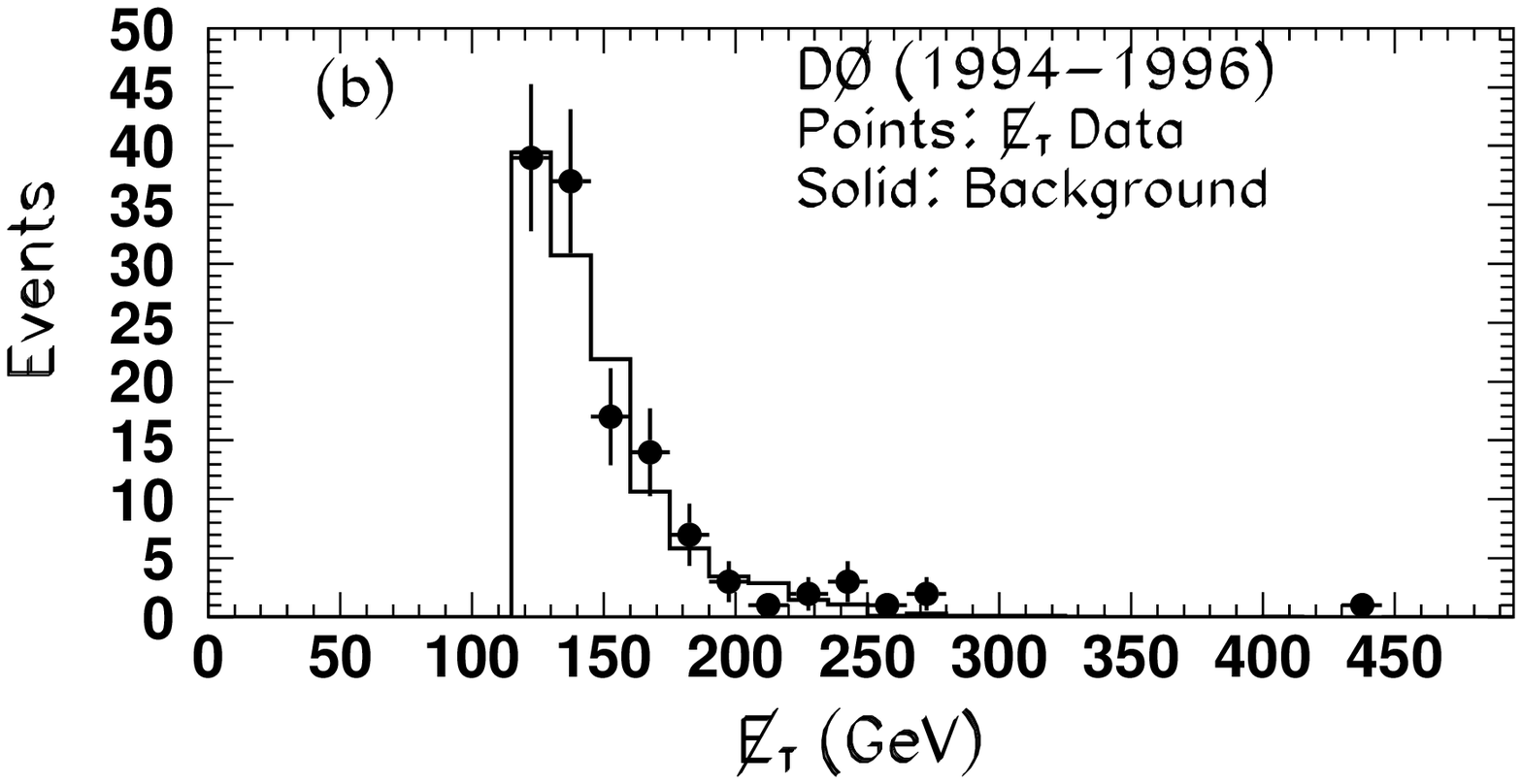}
	}
\vspace{-10pt}
  \caption{(a) Comparison of data (points with error bars) with expected background (dashed histogram), and the sum of background and signal for $n$ = 2, $M_D$ = 800~GeV (dotted histogram), as a function of $\MET$, for $E_T(j_1) > 150$ GeV, $\MET > 150$ GeV, and $E_T(j_2) < 50$ GeV; and (b) comparison of $\MET$ for data with $E_T(j_1) > 115$ GeV, $\MET > 115$ GeV, and $E_T(j_2) < 40$~GeV with expected background.}
\label{fig:fig1}
}  
  \vspace{-10pt}
  \end{figure}

The SM background from $W$ and $Z$-boson production is also modeled by {\sc PYTHIA}, followed by {\sc QSIM} detector simulation. We normalize the $W$ and $Z$ production cross sections to the published D\O\ measurements in the electron channel \cite{crosssection}. The sources of background are detailed in Table \ref{tb:tb2}. With our event selection, the contribution from backgrounds other than $Z(\nu\bar{\nu})~+$ jets is small, and the background from all $W$ and $Z$ sources is estimated to be $30.2 \pm 6.4$ events. The dominant uncertainty on the estimate of $Z(\nu\bar{\nu})~+$ jets is from the uncertainty of the jet-energy scale. The residual background from mismeasured multijet events and cosmic muons is estimated from data, using the uncorrelated $\Delta z$ and $\Delta\phi$ variables described above: we define four data samples, depending on whether the events pass or fail the above criteria; we then normalize the events that fail event vertex confirmation to the candidate sample, using the ratio of the number of events in the two data samples with $\Delta\phi(j_2, \MET) \le 15^{\circ}$; the background from QCD and cosmic rays in the candidate sample is thereby estimated as: 
$$ N_{\rm QCD~+~cosmics} = N^{\Delta z > 10}_{\Delta\phi > 15^\circ} \times N^{\Delta z \le 10}_{\Delta\phi \le 15^\circ}/N^{\Delta z > 10}_{\Delta\phi \le 15^\circ}. $$ This yields $7.8 \pm 7.1$ events. The uncertainty is due primarily to the low statistics of the data samples. The total background estimate is $38 \pm 10$ events.
\begin{table}[tb] \centering
\caption
{The expected and observed number of events in the final jet + $\MET$ sample.
\label{tb:tb2}}
\begin{tabular}
[c]{|c|c|}\hline
Background & $N$ \\\hline
$Z(\nu\nu)+$ jets & 21.0 $\pm$ 5.1 \\
$Z(ee)+$ jets & $<$ 0.01 \\
$Z(\mu\mu)+$ jets & 0.01 $\pm$ 0.01 \\
$Z(\tau\tau)$ (+ jets) & $<$ 0.09 \\
$W(e\nu)+$ jets & 3.1 $\pm$ 0.7 \\
$W(\mu\nu)+$ jets & 0.8 $\pm$ 0.3 \\
$W(\tau\nu)$ (+ jets) & 5.2 $\pm$ 2.3 \\
QCD and cosmics & 7.8 $\pm$ 7.1 \\\hline
Total background & 38.0 $\pm$ 9.6 \\\hline
Data & 38 \\\hline
\end{tabular}
\vspace{-5pt}
\end{table} 
As shown in Fig.\ \ref{fig:fig1a}, the $\MET$ distribution in the data is consistent with that expected from background. Examination of the event with $\MET$ near 450 GeV reveals that the energy deposited by the jet is concentrated in only three calorimeter layers, typical of Bremsstrahlung from a cosmic muon, rather than from a true jet. Nevertheless, the event is kept in the candidate sample, as it passes all {\it a priori\/} selection criteria. From extrapolation, we expect about 0.2 $\pm$ 0.2 background events for $\MET >$ 300 GeV.  

As a cross check of our background estimate, we define a data sample with less stringent requirements, while maintaining roughly the same $E_T(j_1)/E_T(j_2)$ ratio: $E_T(j_1) > 115$ GeV, $\MET > 115$ GeV, and $E_T(j_2) < 40$ GeV. We estimate the background in this sample using the same techniques as described above. This yields an expectation of 105 $\pm$ 16 $W/Z +~$jets events and 16 $\pm$ 9 QCD and cosmic ray events, consistent with the 127 events observed in this data sample. The $\MET$ distributions for this sample and for the expected background are shown in Fig.\ \ref{fig:fig1b}. 

\begin{figure}[htb]
  \vbox{
  \vspace{-0.45in}
  \centerline{
	\includegraphics[trim=0.75in 2.4in 1.2in 1.1in, width=3in]{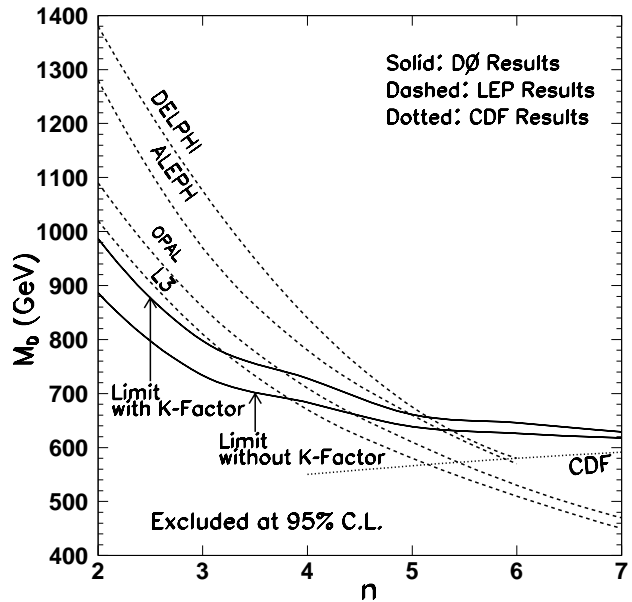}}
  \caption{The 95$\%$ C.L. exclusion contours on the fundamental Planck scale ($M_D$) and number of extra dimensions ($n$) for monojet production at D\O\ (solid lines). The dashed curves correspond to limits from LEP, and the dotted curve is the limit from CDF, both for $\gamma+G_{\rm KK}$ production.}
  \label{fig:fig3}
  }
  \vspace{-25pt}
  \end{figure}

In the absence of evidence for large ED, we calculate upper limits on the cross section for such processes. These limits can be interpreted as lower bounds on the fundamental Planck scale $M_D$ for different integer values of $n$, as listed in Table \ref{tb:tab2}. Using a Bayesian approach \cite{limit}, we set limits on $n$ and $M_D$ using the leading-order cross sections, as well as approximate estimates of next-to-leading-order (NLO) corrections via a constant $K$-factor of 1.34, typical of processes at the Tevatron energies, e.g., Drell-Yan \cite{kfactor1} or direct photon production. As there are no NLO calculations of direct graviton emission to date, the limits with the $K$-factor should be regarded with caution, as purely a measure of sensitivity to the (unknown) NLO effects. The exclusion contours at 95$\%$ confidence, and a comparison with limits from LEP and CDF for the single-photon channel \cite{ledlep,ledcdf}, are shown in Fig.\ \ref{fig:fig3}. While the D\O\ limits are worse than those from LEP at low values of $n$, the sensitivity of the monojet search exceeds the LEP sensitivity at large $n$, due to the higher center-of-mass energy at the Tevatron. The limits correspond to compactification radii ranging from $R < 0.6$~mm ($n = 2$) to $R < 9$~fm ($n = 7$), without correcting for the $K$-factor, and $R < 0.5$~mm ($n = 2$) to $R < 9$~fm ($n = 7$) with approximate NLO effects taken into account. For all $n$, the sensitivity in the single-photon channel at the Tevatron is not as high as in the monojet channel, as the comparison with the CDF limits in Fig.\ \ref{fig:fig3} demonstrates.
\begin{table}[htb] 
\centering
\caption
{95$\%$ C.L. lower limits on $M_D$.}
\label{tb:tab2}
\begin{tabular}
[c]{|c|c|c|c|c|c|c|}\hline
$n$ & 2 & 3 & 4 & 5 & 6 & 7\\\hline
$M_D$ limit without $K$-factor & 0.89 & 0.73 & 0.68 & 0.64 & 0.63 & 0.62\\
scaling (TeV) &  &  &  &  &  & \\\hline
$M_D$ limit with $K$-factor & 0.99 & 0.80 & 0.73 & 0.66 & 0.65 & 0.63\\
scaling (TeV) &  &  &  &  &  & \\\hline
\end{tabular}
\vspace{-5pt}
\end{table}

In summary, we have performed the first search for large extra dimensions in the monojet channel. No evidence for large extra dimensions is observed. We set 95$\%$ confidence-level lower limits on the fundamental Planck scale between 0.6 and 1.0 TeV, depending on the number of extra dimensions. Our limits are complementary to those obtained at LEP in the single photon channel, and are most restrictive to date for $n > 5$. 

% acknowledgement_paragraph_r1.tex            5/02/02
%
We thank Konstantin Matchev for providing us with the {\sc PYTHIA} code for simulation of signal, and for many helpful discussions. We also thank the staffs at Fermilab and collaborating institutions, 
and acknowledge support from the 
Department of Energy and National Science Foundation (USA),  
Commissariat  \` a L'Energie Atomique and 
CNRS/Institut National de Physique Nucl\'eaire et 
de Physique des Particules (France), 
Ministry for Science and Technology and Ministry for Atomic 
   Energy (Russia),
CAPES and CNPq (Brazil),
Departments of Atomic Energy and Science and Education (India),
Colciencias (Colombia),
CONACyT (Mexico),
Ministry of Education and KOSEF (Korea),
CONICET and UBACyT (Argentina),
The Foundation for Fundamental Research on Matter (The Netherlands),
PPARC (United Kingdom),
Ministry of Education (Czech Republic),
A.P.~Sloan Foundation,
and the Research Corporation.
% 
%\begin{references}
%\input{list_of_visitor_addresses_r1.tex}

%\end{references}

\begin{thebibliography}{99}
\bibitem{blackhole} Recent calculations of the production of mini-black-holes in high-energy particle collisions \cite{4a}, applied to results from the HiRes and AGASA experiments \cite{4b}, indicate a sensitivity competitive with that of collider experiments for small values of $n$.
\bibitem {etad}Detector pseudorapidity $\eta_{d}$ is defined as $-\ln\left(\tan\left(\theta_{d}/2\right)\right)$, where $\theta_{d}$ is the polar angle with respect to the proton beam, as measured relative to the center of the detector.
\end{thebibliography}
\end{document}